\documentclass[floats,aps,amsfonts]{revtex4}
\usepackage{multirow}

\usepackage{graphicx}

\def\alt{\mathrel{\hbox{\rlap{\hbox{\lower4pt\hbox{$\sim$}}}\hbox{$<$}}}}
\def\agt{\mathrel{\hbox{\rlap{\hbox{\lower4pt\hbox{$\sim$}}}\hbox{$>$}}}}
\newcommand {\be} {\begin{equation}} \newcommand {\ee}
{\end{equation}} \newcommand {\ban} {\begin{eqnarray}} \newcommand
{\ean} {\end{eqnarray}}

\begin{document}

\title{Using LISA EMRI sources to test off-Kerr deviations in the geometry
of massive black holes }

\author{Leor Barack}
\address
{School of Mathematics, University of Southampton, Southampton,
SO17 1BJ, United Kingdom}
\author{Curt Cutler}
\address
{Jet Propulsion Laboratory, California Institute of Technology, Pasadena,
CA 91109}

\date{\today}

\begin{abstract}

Inspirals of stellar-mass compact objects (COs) into $\sim 10^6 M_{\odot}$
black holes are especially interesting sources of gravitational-waves for
the planned Laser Interferometer Space Antenna (LISA).
The orbits of these extreme-mass-ratio inspirals (EMRIs) are highly
relativistic, displaying extreme versions of both perihelion precession and
Lense-Thirring precession of the orbital plane.  We investigate the question
of whether the emitted  waveforms can be used to strongly constrain the
geometry of the central massive object, and in essence check
that it corresponds to a Kerr black hole (BH).  For a Kerr BH, all multipole
moments of the spacetime have a simple, unique relation to $M$ and $S$, the
BH mass and spin;  in particular, the spacetime's  mass quadrupole moment
$Q$ is given by $Q = - S^2/M$.  Here we treat $Q$ as an additional parameter,
independent of $S$ and $M$, and ask how well observation can constrain its
difference from the Kerr value.  This was already estimated by Ryan, but for
the simplified case of circular, equatorial orbits, and Ryan also neglected
the signal modulations arising from the motion of the LISA satellites.
We consider generic orbits and include the modulations due to the satellite motions.
For this analysis, we use a family of approximate (basically post-Newtonian)
waveforms, which represent the full parameter space of EMRI sources, and which
exhibit the main qualitative features of true, general relativistic waveforms.
We extend this parameter space to include (in an approximate manner) an arbitrary
value of $Q$, and then construct the Fisher information matrix for the extended
parameter space. By inverting the Fisher matrix, we estimate how accurately
$Q$ could be extracted from LISA observations of EMRIs.
For 1 year of coherent data from the inspiral of a $10 M_{\odot}$ black hole into
rotating black holes of masses
$10^{5.5} M_{\odot}$, $10^6  M_{\odot}$, or $10^{6.5} M_{\odot}$, we find
$\Delta (Q/M^3) \sim 10^{-4}$, $10^{-3}$, or $10^{-2}$, respectively (assuming total
signal-to-noise ratio of 100, typical of the brightest detectable EMRIs). These results
depend only weakly on the eccentricity of the inspiral orbit or the spin of
the central object.


\end{abstract}

\maketitle

\section{Introduction}

Inspirals of stellar-mass compact objects (COs) captured by massive ($\sim 10^6 M_{\odot}$)
black holes (MBHs) in  galactic nuclei are especially interesting gravitational-wave sources for
the planned Laser Interferometer Space Antenna (LISA).  Because the ratio of CO mass $\mu$ to
MBH mass $M$ is typically $\mu/M \sim 10^{-5}$, these events are generally referred to
as extreme-mass-ratio inspirals (EMRIs). Current estimates are that LISA will detect of
order $\sim 10^2$ EMRIs per year~\cite{hopman_alexander,miller,steinn,EMRIs}.
Captures of $\sim 10 M_{\odot}$
black holes (BHs) are expected to dominate the detection rate (although some
captured white dwarfs and neutron stars  will be detected as well), both because
mass segregation in galactic nuclei tends to concentrate BHs very near the
central MBH~\cite{hopman_alexander}, and because the inspiraling $\sim 10 M_{\odot}$
black holes are intrinsically stronger sources than the white dwarfs
or neutron stars, and so are detectable to a much greater
distance: out to $z \sim 1$.

The orbits of the inspiralling COs are highly
relativistic,  displaying extreme versions of both perihelion precession and Lense-Thirring
precession of the orbital plane. Because of the tiny mass ratio, the inspiral is quite slow;
individual inspiral waveforms will be observable for years, or equivalently
for $\gtrsim 10^5$ cycles.  While at any instant the EMRI waveforms will be buried in both
LISA's instrumental noise and a gravitational-wave foreground from Galactic binaries,
it will nevertheless be possible to dig out the EMRI signals  using techniques based on
matched filtering~\cite{EMRIs}, thanks to this large number of cycles.
The matched-filtering signal-to-noise ratio (SNR), which
scales like the square root of the number of observed cycles, will typically be in the
range $\sim 30-300$ for EMRI  detections.
Because of this large SNR, and because the  complex waveforms
are quite sensitive to the source's physical parameters, it will be possible to extract these
source parameters with high accuracy; in particular,
it will  be possible to infer the mass of the CO, and the
mass and spin of the MBH, all to within fractional accuracy
$\sim 10^{-4}$~\cite{BC1}.

The famous ``no hair'' theorem of General Relativity
essentially states that, almost immediately after its formation, a BH ``settles down'' to one
of the Kerr solutions, and so its entire geometry is  characterized by just two
physical parameters:
its mass $M$ and its spin angular momentum $S$.
In particular, all the multipole moments of a Kerr spacetime can be
expressed in terms of $M$ and $S$ alone, 
according to the relation
\begin{equation}\label{moments}
M_l+iS_l=M(ia)^l,
\end{equation}
where $M_l$ and $S_l$ are the mass and mass-current multipole moments, respectively,
and $a=S/M$ is the standard Kerr spin parameter.
For example, the mass quadrupole moment $Q \equiv M_2$ of the pure Kerr geometry is given by
\begin{equation}\label{KerrQ}
Q = - S^2/M.
\end{equation}

It has been suggested by several authors (e.g., \cite{ryan95,ryan97I,Hproc}) that LISA
detections of EMRIs could provide a highly accurate observational test of the ``Kerr-ness''
of the central massive object. (In what follows we will continue to refer to the central
massive object as ``MBH'' for brevity, even though in our analysis we allow it to be
a non-black hole object.)  Indeed, Ryan \cite{ryan95} showed
that for inspiralling trajectories that are slightly eccentric and slightly non-equatorial,
in principle all multipole moments of the spacetimes are redundantly encoded in the
emitted gravitational waves, through the time-evolution of the orbit's three fundamental
frequencies. (Basically, these are the fundamental frequencies associated with the $\phi$, $r$ and
$\theta$ motions; see \cite{Schmidt,drasco_hughes}
for a proof that the waveform corresponding to a Kerr geodesic is a discrete sum of
integer harmonics of these $3$ fundamental frequencies.)

%
%
Ryan~\cite{ryan97I} went on to estimate how accurately the central massive
object's lower-order multipole moments could be measured by LISA.
Because the CO's complex motion depends strongly on the background
geometry, one might expect to be able to measure the lowest moments rather well. Basically,
the long-term phase evolution of the orbit (and that of the emitted
gravitational wave) can act as a microscope, probing the fine features
of the geometry.
Lacking a theory that accurately describes the orbital
evolution in spacetimes with arbitrary multipole structure (in fact, such a theoretical
framework is not yet at hand, even for the pure Kerr case), Ryan considered a simple
waveform model based on lower-order post-Newtonian theory. For
simplicity, he also restricted the EMRI parameter space to orbits that are
circular and
equatorial, and he neglected the modulations imposed
on the observed signal by the spacecraft motions.  Within these simplifications,
he calculated the Fisher matrix  on the space of EMRI waveforms, and inverted it
to obtain (to lowest order in SNR$^{-1}$) the variance-covariance matrix, which
describes the distribution of statistical errors associated with a given measurement.
Ryan found, for example,
that for a capture of a $10 M_{\odot}$ black hole by a
$10^6 M_{\odot}$ central object,
assuming 2 years of inspiral data with
SNR of 100, LISA could measure $Q/M^3$ to within $\sim 5\times 10^{-1}$, $M$ to
within a fractional error of $\sim 3\times 10^{-4}$, $S/M^2$ to within $\sim 10^{-2}$, and
$\mu$ to within $\sim 10^{-3}$.
(Ryan~\cite{ryan97I} further showed how these accuracies get degraded as additional multipole
moments, besides $Q \equiv M_2$, get treated as independent quantities in the signal model.)

In essence,  our paper improves on Ryan's 1997 analysis~\cite{ryan97I} by considering the case
of generic (i.e., non-circular, non-equatorial) orbits,  and by employing a much
more realistic treatment of LISA's instrumental response, including the Doppler modulation
due to the constellation's center-of-mass motion and the additional modulations due
to the rotational motion of the individual satellites with respect to that center-of-mass.
Our analysis is based on our previously introduced family of ``analytic kludge''
(basically post-Newtonian) waveforms~\cite{BC1}, representing practically the full
parameter space of capture sources (though neglecting the spin of the CO, for reasons described below).
We extend this parameter space to include, in an approximate manner, an
arbitrary deviation in the value of the central object's quadrupole moment, and
we then construct the Fisher information matrix for the extended parameter space.
By inverting the Fisher matrix, we estimate how accurately the central object's quadrupole moment
$Q$ can be determined, independently of its mass and spin.

A priori, it would seem difficult to predict how our more realistic treatment should affect  Ryan's results.
On the one hand, our more realistic and complicated waveforms clearly
encode more information.  On the other hand, our waveform depends on all $15$ physical parameters (including $Q$), while
Ryan's more simplistic signal model was based on
only $7$ parameters~\footnote{Our additional $8$ parameters are: $2$ sky-position angles, $2$ angles describing the
MBH's spin direction, $2$ angles describing the initial orientation of the orbital angular momentum, $1$ angle describing
the direction to pericenter, and the final value of the eccentricity, $e_{\rm LSO}$.}; of course,
one's ability to extract any given source parameter
(such as $Q$) is diluted when more parameters are
added to the source description.  What we find, in the end, is that $Q/M^3$ can be
measured to $\sim 10^{-4}-10^{-2}$ (based on the last year of data and assuming SNR$=100$).
Thus EMRIs provide an even better test of ``Kerr-ness'' than Ryan estimated.


Again, the idea of the Kerr-ness test explored here is to treat $M, S$, and $Q$ as three independent quantities, and
then see whether the LISA data confirms the Kerr relation: $Q = -S^2/M$.   This test is similar in spirit to the
well-known tests of general relativity made with the binary pulsar PSR 1913+16 (see, e.g., ~\cite{lorimer}).
The binary's Keplerian parameters (like its period
and eccentricity) and it post-Keplerian parameters (e.g., the precession rate $\dot \omega$, Einstein delay $\gamma$,
and period derivative $\dot P_b$) are all fit for independently.  Within General Relativity these are
redundant, since all parameters {\it excluding} $\dot P_b$ lead to a unique physical description of the binary.
The test is then to see whether the measured $\dot P_b$ is indeed the value that GR predicts.
Our EMRI test treats $Q$ in an analogous way to $\dot P_b$ for PSR 1913+16.
Of course, testing that the spacetime's quadrupole moment coincides with the Kerr value is only
one possible test of ``Kerr-ness'', and not necessarily the best one.  But absent a compelling alternative to
the ``null hypothesis'' that the MBH is a Kerr BH, tests of this sort are perhaps the best one can do.

We briefly mention some recent related work.  In \cite{collins_hughes}, Collins and Hughes
began a research program aimed at extending Ryan's work, beginning with the construction of
exact spacetime solutions corresponding to ``bumpy'' black holes.
Also recently, Glampedakis and Babak \cite{gb}  analyzed geodesic orbits in
both Kerr and slightly non-Kerr spacetimes, and concluded that any
effects due to non-Kerr multipole moments would be difficult to detect, since
they would be mimicked by small, compensating offsets in the geodesic's initial conditions.
Glampedakis and Babak \cite{gb} did {\it not} consider the effects of radiation
reaction---i.e., they effectively considered only short stretches of actual
waveforms---so the implications of their work for actual, {\it inspiralling} trajectories
remained unclear. (Work to include radiation reaction in this model is now in
progress \cite{Kostas private}).
Finally, Kesden, Gair, and Kamionkowski~\cite{boson} have investigated the question of
whether EMRI waveforms can be used to distinguish between a central black hole and a
massive boson star.  They argue persuasively that one could easily distinguish between
these two types of central object, since in the Kerr case the waveform ``shuts off''
shortly after the CO reaches the last stable orbit (LSO), while in the boson star case the inspiral
continues until the CO comes to rest at the center of the boson star.

\section{``Analytic kludge'' EMRI waveforms}
\subsection{Overview}
Since highly accurate inspiral waveforms for EMRIs are not yet available
(due to the difficulty of solving the radiation reaction problem for generic
orbits in Kerr \cite{PoissonRev}), approximate, or ``kludged'', families of waveforms
have been developed for use in EMRI investigations until more exact versions are available.
One such family is the ``numerical kludge'' waveforms developed by Gair, Glampedakis and
collaborators \cite{GK,GG,Num kludge}, and another is the
``analytic kludge'' waveforms developed by Barack and Cutler (hereinafter BC~\cite{BC1}).
Our ``analytic kludge'' waveforms are less accurate than the ``numerical kludge'' ones,
but they are simpler to compute and more tractable for the purpose of accurately
calculating partial derivatives (with respect to source parameters) numerically,
as required for producing the Fisher matrix. The BC waveforms were used for just
this purpose in Ref.~\cite{BC1}, where it was estimated that LISA could measure
the masses of both bodies and the spin magnitude of the massive black
hole to fractional accuracy $\sim 10^{-5}-10^{-4}$ (for SNR =100).  In this paper we modify the
analytic kludge evolution equations to include the influence of
possible deviations in the quadrupole moment of the MBH.
We now summarize this ``Q-enhanced'' version of our waveform family;
we refer the reader to Ref.\ \cite{BC1} for more details.

In the ``analytic kludge'' approach, we approximate the CO-MBH system as being, at any instant,
a Newtonian-orbit binary emitting a Peters-Matthews \cite{pm} (i.e., lowest order,
quadrupolar) waveform. We then use post-Newtonian (PN) equations to secularly evolve the parameters
of the orbit. In particular, we include orbital decay from radiation reaction,
pericenter precession, and Lense-Thirring precession of the orbital plane.
%
The motion of the LISA detector introduces additional modulations; for these
we adopt the low-frequency approximation developed by  Cutler~\cite{cutler98}.
Cutler's treatment is not very accurate for frequencies
$f \gtrsim 30$ mHz (where the gravitational-wave period is longer than the
round-trip travel time for light up and down the LISA arms),
but it is adequate for our EMRI analysis, since
most of the EMRI SNR accumulates in the range $\sim 1 - 10$ mHz.
Since the main physical effects are all included, and their dependence on the
source parameters are given correctly to (at least) the lowest non-trivial post-Newtonian order,
we expect that parameter estimation accuracies for our ``kludge'' waveforms are a reasonable guide
to the accuracies we can expect (i.e., are probably correct to within an order of magnitude), once the true,
general relativistic waveforms are in hand.  We test this expectation to some degree in
Sec.\ III.C, where we simply remove (by hand) some $P^1N$ terms from our kludged evolution
equations, and find that this modifies the deduced measurement error in $Q/M^3$ by
a factor of less than $3$.
%
%

\subsection{Parameter Space}

Assuming the MBH is a Kerr back hole, the two-body system is
completely described by 17 parameters. However, the spin of the
CO  is at most marginally relevant~\cite{BC1}, so, as in BC, we shall neglect it in our
analysis.~\footnote{More specifically, it was shown in Appendix C of \cite{BC1} that
effects related to the CO's spin can modify the waveform phase
by at most a few radians out of $\sim 10^6$, for a maximally spinning CO.
This is far smaller than the sought-for influence of an excess quadrupole moment
for the central object: A quadrupole moment twice its Kerr value would modify
the year-long inspiral waveform by $\sim 10^3$ radians.}
That leaves 14 parameters, assuming a massive Kerr BH. We now include one additional
parameter, describing the quadrupole moment of the geometry. More precisely, we
choose as our additional parameter the dimensionless quantity
$$
\tilde Q  \equiv Q/M^3.
$$


%
%
Including $\tilde Q$, we therefore have a 15-dimensional parameter space.
For completeness, we list the full parameter set here.
This list is identical to the one in BC, except for the inclusion of $\tilde Q$
(and a re-definition of $t_0$---see below).
\begin{eqnarray} \label{lambda}
\lambda^a &\equiv& (\lambda^0,\ldots,\lambda^{14}) =\nonumber\\
&&
\left[t_0(\times 1\;{\rm mHz}),\,\ln\mu,\,\ln M,\,\tilde S,\,e_{\rm LSO},\,\tilde\gamma_0,\,\Phi_0,\,
\mu_S\equiv\cos\theta_S,\,\phi_S,\,\cos\lambda,\,\alpha_0,
\mu_K\equiv\cos\theta_K,\,\phi_K,\,\ln(\mu/D), \tilde{Q} \right].
\end{eqnarray}
Here, $t_0$ specifies the instant of time where the CO reaches the LSO,
and the inspiral transits to a plunge. $\mu$ and $M$ are the masses of the CO and MBH,
respectively, and $\tilde S=S/M^2$, where $S$ is the magnitude of the
MBH's spin angular momentum.
The parameters $e_{\rm LSO}$, $\tilde\gamma_0$, and $\Phi_0$ describe,
respectively, the eccentricity, the direction of the pericenter within
the orbital plane, and the mean anomaly---all at time $t_0$ (time of plunge).
More specifically, we take
$\Phi_0$ to be the mean anomaly with respect to pericenter passage, and
$\tilde\gamma_0$ to be the angle (in the plane of the orbit) from
$\hat L \times \hat S$ to pericenter (where $\hat L$ is a unit vector in the
direction of the orbital angular momentum).
The parameter $\alpha_0\equiv\alpha(t=t_0)$ describes the direction
of $\hat L$ around $\hat S$ at time $t_0$.
[See Eq.~(18) of BC for the precise definition of $\alpha(t)$.]
The angles $(\theta_S,\phi_S)$ describe the direction to the source, in
ecliptic-based coordinates; $(\theta_K,\,\phi_K)$ represent the direction
$\hat S$ of the MBH's spin (approximated as constant) in ecliptic-based
coordinates; and $\lambda$ is the angle between $\hat L$ and $\hat S$ (also
approximated as constant).
Finally, $D$ is the distance to the source, and, again, $\tilde{Q}\equiv Q/M^3$
is spacetime's dimensionless quadrupole moment.
The various parameters and their meaning are summarized in Table
\ref{tableI}.  Fig.~1 of BC illustrates the various angles involved
in our parameterization.

Note that for simplicity we are treating the background spacetime as
Minkowski space, not Robertson-Walker. To correct this, for a source
at redshift $z$, requires only
the simple translation: $M\rightarrow M(1+z)$, $\mu\rightarrow \mu(1+z)$,
$S\rightarrow S(1+z)^2$, $D\rightarrow D_L$, where $D_L$ is the
``luminosity distance''~\cite{markovic}.

\begin{table}[thb]
\centerline{$\begin{array}{c|c|l}\hline\hline
\lambda^0 & t_0(\times 1 {\rm mHz})   & \text{time when CO reaches LSO}   \\
\lambda^1 & \ln\mu        & \text {($\ln$ of) CO's mass}\\
\lambda^2 & \ln M         & \text {($\ln$ of) MBH's mass}\\
\lambda^3 & \tilde S\equiv S/M^2  & \text{dimensionless magnitude of (specific)
                         spin angular momentum of MBH} \\
\lambda^4 & e_{\rm LSO}          & \text{eccentricity at $t_0$ (i.e., final eccentricity)} \\
\lambda^5 & \tilde\gamma_0 &  \text{$\tilde\gamma(t_0)$,
    where $\tilde\gamma(t)$ is the angle (in orbital plane)
    between $\hat L\times\hat S$ and pericenter}      \\
\lambda^6 & \Phi_0        & \text{$\Phi(t_0)$, where $\Phi(t)$ is
    the mean anomaly}\\
\lambda^7 & \mu_S\equiv\cos\theta_S  & \text{(cosine of) the source direction's
    polar angle }  \\
\lambda^8 & \phi_S        & \text{azimuthal direction to source}  \\
\lambda^9 & \cos\lambda   & \hat L\cdot\hat S(={\rm const}) \\
\lambda^{10} & \alpha_0     & \text{$\alpha(t_0)$, where $\alpha(t)$
    is the azimuthal direction of $\hat L$ (in the orbital plane)}   \\
\lambda^{11} & \mu_K\equiv\cos\theta_K & \text{(cosine of) the polar angle
   of MBH's spin}  \\
\lambda^{12} & \phi_K       &  \text{azimuthal direction of MBH's
    spin}  \\
\lambda^{13} & \ln(\mu/D)          & \text{($\ln$ of) CO's mass divided by distance to source}\\
\lambda^{14} & \tilde{Q}\equiv Q/M^3 & \text{dimensionless quadrupole moment of MBH}\\
\hline\hline
\end{array}$}
\caption{\protect\footnotesize
Summary of physical parameters and their meaning.
For further details see Fig.~1 of BC and the description
in the text.}\label{tableI}
\end{table}

\subsection{Orbital evolution with arbitrary quadrupole moment}

The inspiralling orbit is determined by the evolution equations for $\Phi(t), \nu(t), \tilde \gamma(t),
e(t), \alpha(t)$.  (The 2 masses, the MBH spin vector $\vec S$, the angle $\lambda$ and
of course the source's sky position are all taken to be constants.)
The post-Newtonian evolution equations we adopt are:

\begin{eqnarray}
\frac{d\Phi}{dt} &=& 2\pi\nu, \label{Phidot} \\
%
\left(\frac{d\nu}{dt}\right) &=&
\frac{96}{10\pi}(\mu/M^3)(2\pi M\nu)^{11/3}(1-e^2)^{-9/2}
\bigl\{
\left[1+(73/24)e^2+(37/96)e^4\right](1-e^2) \label{nudotKerr}\nonumber \\
&&+ (2\pi M\nu)^{2/3}\left[(1273/336)-(2561/224)e^2-(3885/128)e^4
-(13147/5376)e^6 \right] \nonumber \\
&&- (2\pi M\nu)\tilde S\cos\lambda (1-e^2)^{-1/2}\bigl[(73/12)
+ (1211/24)e^2
+(3143/96)e^4 +(65/64)e^6 \bigr] \nonumber\\
&&-(2\pi M\nu)^{4/3}\tilde{Q}(1-e^2)^{-1}
\left[(33/16)+(359/32)e^2-(527/96)\sin^2\lambda\right]
\bigr\}, \label{nudot} \\
%
\left(\frac{d\tilde\gamma}{dt}\right) &=& 6\pi\nu(2\pi\nu M)^{2/3} (1-e^2)^{-1}
\left[1+\frac{1}{4}(2\pi\nu M)^{2/3} (1-e^2)^{-1}(26-15e^2)\right] \nonumber \\
&&-12\pi\nu\cos\lambda \tilde S (2\pi M\nu)(1-e^2)^{-3/2} \nonumber\\
&& -\frac{3}{2}\pi\nu \tilde{Q}(2\pi M\nu)^{4/3}(1-e^2)^{-2}\left(5\cos\lambda-1\right),
\label{Gamdot} \\
%
\left(\frac{de}{dt}\right)  &=& -\frac{e}{15}(\mu/M^2) (1-e^2)^{-7/2} (2\pi M\nu)^{8/3}
\bigl[(304+121e^2)(1-e^2)\bigl(1 + 12 (2\pi M\nu)^{2/3}\bigr) \, \nonumber \\
&&- \frac{1}{56}(2\pi M\nu)^{2/3}\bigl( (8)(16705) + (12)(9082)e^2 - 25211e^4 \bigr)\bigr]\,
\nonumber \\
&&+ e (\mu/M^2)\tilde S\cos\lambda\,(2\pi M\nu)^{11/3}(1-e^2)^{-4}
\, \bigl[(1364/5) + (5032/15)e^2 + (263/10)e^4\bigr] ,
\label{edot} \\
%
\left(\frac{d\alpha}{dt}\right)&=& 4\pi\nu \tilde S (2\pi M\nu)(1-e^2)^{-3/2}
\nonumber\\
&&+3\pi\nu\tilde{Q}(2\pi M\nu)^{4/3}(1-e^2)^{-2}\cos\lambda.
\label{alphadot}
\end{eqnarray}
These are the same evolution equations as used in BC, except that the lowest-order terms
$\propto \tilde Q$ have been added to the right-hand side of Eqs.~(\ref{nudot}), (\ref{Gamdot}), and
(\ref{alphadot}).
For the two precession rates, ${d\tilde\gamma}/{dt}$ and ${d\alpha}/{dt}$, these
extra terms represent essentially Newtonian effects: the Newtonian precession rates for orbits around
oblate bodies.
The term $\propto\tilde Q$ in our Eq.~(\ref{Gamdot}) is taken from Eq.~(3) of Lai {\it et al}.~\cite{lai}, using
the identifications $\chi \rightarrow \tilde\gamma$, $I_1 - I_3 \rightarrow Q$ and $\theta \rightarrow \lambda$, and
the approximation
$\theta \approx \theta_c$ for $\mu/M \ll 1$.
The term $\propto\tilde Q$  in Eq.~(\ref{alphadot})  is taken from Eq.~(4) of Lai {\it et al}.~\cite{lai}.
To obtain the term $\propto\tilde Q$
in Eq.~(\ref{nudot})  for $\dot\nu$,
we start with the Newtonian relation
\begin{equation}
\nu = \frac{1}{2\pi M}(-2 E /\mu)^{3/2}.
\end{equation}
This relation is {\it not} modified by $Q$, through first order [cf. Eq. (2.26) of
Gergely et al.~\cite{gergely_03} or Eqs.~(A.11)-(A.12) of Ashby~\cite{ashby}].
Therefore we can write, through first order in $Q$ and to lowest non-trivial order in a
PN expansion:
\begin{equation}\label{nudot-Q}
\dot \nu_Q = \frac{d\nu}{dE}\dot E_Q = -\frac{3}{\mu} \nu (2\pi M\nu)^{-2/3}{\dot E}_Q \, .
\end{equation}
where the subscript `Q' here means ``the piece linear in Q''.
For $\dot E_Q$, we adopt the approximate formula given in
Eq.(44) of Gair and Glampedakis~\cite{GG}. [As those authors note, their Eq.\ (44)
is unfortunately ``missing'' a piece proportional to $e^2 \sin^2\lambda$,
since the coefficient of that term has not yet been derived.]
Using the identifications that their $q^2$ corresponds to our
$-\tilde Q$ and their $\iota$  to our $\lambda$, and then plugging $\dot E_Q$ into
Eq.~(\ref{nudot-Q}),
we obtain the
term  $\propto \tilde Q$ in  Eq.~(\ref{nudot}).

For consistency, our Eq.~(\ref{edot}) should also contain a term  $\propto\tilde Q$; however
it would be nontrivial to derive this term from the existing literature, and its influence
on astrophysically realistic waveforms is probably substantially less than the terms
$\propto\tilde Q$ that we {\it have} included, since $e^2 \lesssim 0.1$ for typical inspirals.
It is a feature of our
kludged waveforms that we attempt to include the most important terms but do not worry about
consistently including all terms out to some fixed order; in this spirit, we simply neglect
the Q-correction to $\dot e$.

%
%
%

To construct the orbit, we start by choosing some final eccentricity $e_{\rm LSO}$.
In practice we consider moderate final eccentricities: those in the range
$0\leq e_{\rm LSO}\leq 0.3$. We then set the final radial frequency as
\be \label{nu_LSO}
\nu_{\rm LSO} = 0.0415\times(2\pi M)^{-1},
\ee
which reproduces the correct gravitational-wave frequency at the LSO through Eq.\
(\ref{GWfreq}) (explained further below).
This LSO value of the frequency
applies to circular orbits in Schwarzschild ($S=e=0$), but for simplicity we shall
use it here for any of the moderate eccentricities considered, and for any spin.
We then integrate the evolution equations, Eqs.~(4)--(8), backwards in time,
starting from the LSO. Given the solution of these ordinary differential equations,
it is straightforward to construct both polarizations of the emitted waveform using
Eqs.\ (7)--(10) in BC:
\begin{equation}\label{polarizations}
h_{+,\times}(t) = \sum_n \big[ A_{n,s}^{+,\times}{\rm sin}\Phi_n(t) +
A_{n,c}^{+,\times}{\rm cos}\Phi_n(t) \big] \, ,
\end{equation}
where the subscript ``n'' labels the different harmonics of the orbital frequency,
and the subscripts ``s'' and ``c'' just refer to the two quadratures, sine and cosine.
In practice, we find it sufficient, given the range of LSO eccentricities considered
in this work, to include in the above sum only the first $20$ harmonics.
Explicit formulae for the $A_{n,s}^{+,\times}$ and $A_{n,c}^{+,\times}$ are trivially obtained from
Eqs.~(7), (8), and (10) of BC, and the
phase of the $n^{th}$ harmonic, $\Phi_n(t)$, is given by
\begin{equation}\label{phase}
\Phi_n(t) = n\Phi_0 + 2\pi \int_{t_0}^t f_n(t') dt' +  2\pi f_n(t) \hat n \cdot \vec R(t) \, .
\end{equation}
Here $f_n(t')$ is the instantaneous frequency of the $n^{th}$ harmonic, and the second term
in Eq.~(\ref{phase}) is the Doppler delay, where $\hat n$ is the direction to the source and
$\vec R(t)$ is the position of LISA's center with respect to the Solar System
Barycenter~\footnote{Eq.~(33) of BC incorrectly gives the Doppler piece of
$\Phi_n(t)$ as $2\pi n \nu(t)  \hat n \cdot \vec R(t)$. However the
code that generated the results in BC used the same approximation as given
in this paper: $2\pi f_n(t) \hat n \cdot \vec R(t)$.}.
For our ``kludge'' waveforms, we approximate
$f_n(t')$ by simply~
\begin{equation} \label{GWfreq}
f_n(t') = n \, \nu(t') + \dot{\tilde\gamma}(t')/\pi,
\end{equation}
the motivation for which is explained in the last paragraph of Section IV of BC.

To explain our choice for $\nu_{\rm LSO}$ [Eq.\ (\ref{nu_LSO}) above], recall
that, for circular orbits around a Schwarzschild black hole, the (azimuthal) orbital
frequency at the LSO is precisely $\nu_{\rm LSO}=6^{-3/2}(2\pi M)^{-1}$, with
a corresponding gravitational wave frequency of $f_{\rm GW}=2\times 6^{-3/2}(2\pi M)^{-1}$
(since the radiation in this case is dominated by the $n=2$ harmonic).
To be able to reproduce this value of $f_{\rm GW}$ at the LSO through Eq.\
(\ref{GWfreq}), we must require
$2\nu_{\rm LSO}+\dot{\tilde\gamma}_{\rm LSO}/\pi=2\times 6^{-3/2}(2\pi M)^{-1}$,
where $\dot{\tilde\gamma}_{\rm LSO}$ is obtained from Eq.\ (\ref{Gamdot}) by
replacing $\nu\to \nu_{\rm LSO}$ and neglecting the terms proportional to $e^2$,
$\tilde S$, and $\tilde Q$.
Solving this condition for $\nu_{\rm LSO}$ gives Eq.\ (\ref{nu_LSO}) above.

In the frequency range of interest, we can regard LISA's output as equivalent to two independent
data streams, $s_I(t)$ and $s_{II}(t)$. We idealize each of these as being the sum of
(stationary, Gaussian) noise $n_{I,II}(t)$ and a gravitational-wave signal, $h_{I,II}(t)$:
\be
s_I(t) = n_I(t) + h_I(t), \ \ \ s_{II}(t) = n_{II}(t) + h_{II}(t) \, .
\ee
The gravitational-wave signals in the two channels,  $h_I(t)$ and $h_{II}(t)$,
are related to the two incoming polarizations $h_{+,\times}(t)$
by time-varying transfer
functions, $F_{I}^{+,\times}(t)$ and $F_{II}^{+,\times}(t)$.
Thus we have
\begin{equation}\label{transfer}
h_I(t)
=\frac{\sqrt{3}}{2}\big[ F_{I}^+(t) h_{+}(t) + F_{I}^{\times}(t) h_{\times}(t)\big]  \, ,
\end{equation}
and similarly for $h_{II}(t)$.
The equations used to construct these transfer functions are given explicitly in BC.
(The factor $\sqrt{3}/2$ is just the sine of $60^{\circ}$, the angle between the LISA arms; it could have been included
in the definition of the transfer functions, but it was left as a separate factor so that the $F$'s
would coincide with LIGO transfer functions.)

We emphasize  that in our ``analytic kludge'' scheme, the value of the quadrupole moment
affects the inspiral trajectory (i.e., it modifies the sequence of Keplerian ellipses that
the orbit osculates though), but $\tilde Q$ does {\em not} enter into the algorithm for
constructing the gravitational waveform from that trajectory; i.e.,
we continue to use the quadrupole formula to construct the waveform from the inspiral trajectory.
Since the quadrupole formula does a surprisingly good job of reproducing waveforms for Kerr~\cite{Num kludge}, we expect
this also to be the case for spacetimes that are slightly off-Kerr.


Finally, we briefly review the signal analysis formalism necessary for calculating $\Delta\tilde Q$
(the measurement accuracy in $\tilde Q$) in order to make clear our approximations;
we refer to BC for more details.  The (sky-averaged) noise spectral density $S_h(f)$
for each of the two channels, I and II, determines a natural inner product on the space of LISA data
streams, as follows.  Let  ${\bf p}(t) \equiv [p_I(t) , p_{II}(t)]$
and ${\bf q}(t) \equiv [q_I(t) , q_{II}(t)]$ be two different LISA data sets.
Their inner product is~\cite{cutler_flanagan}:
\begin{equation}\label{inner}
\left( {\bf p} \,|\, {\bf q} \right)
\equiv 2\sum_{\alpha=I}^{II} \int_0^{\infty}\left[ \tilde p_\alpha^*(f)
\tilde q_\alpha(f) + \tilde p_\alpha(f) \tilde q_\alpha^*(f)\right]
/\left[\frac{3}{20}S_h(f)\right]\,df \, .
\end{equation}
[The factor $\frac{3}{20}$ is due to the ``sky-averaging'' convention adopted here in
the definition of $S_h(f)$.]
This inner product has the property that the matched-filtering SNR for
any imbedded gravitational waveform ${\bf h}(t) \equiv [h_I(t) , h_{II}(t)]$
is just its norm:
\be
{\rm SNR}[{\bf h}] =  \left( {\bf h} \,|\, {\bf h} \right)^{1/2} \, .
\ee

The Fisher information matrix $\Gamma_{ab}$ is defined by
\begin{equation}
\label{sig}
\Gamma_{ab} \equiv \bigg( {\partial {\bf h} \over \partial \lambda^a}\, \bigg| \,
{\partial {\bf h} \over \partial \lambda^b }\bigg),
\end{equation}
where $a,b,\ldots$ are parameter-space indices.
To lowest order in an expansion in SNR$^{-1}$, the variance-covariance matrix
for the errors is just the inverse of the Fisher matrix:
\be
\langle \delta \lambda^a  \delta\lambda^b \rangle
= \big(\Gamma^{-1})^{ab} \big[1 + {\cal O}({\rm SNR}^{-1})\big],
\ee
where ``$\langle \cdots  \rangle$'' means ``expectation value''.
Hence the measurement error in the various parameters is given by
\be
\Delta \lambda^a \equiv \langle \delta \lambda^a  \delta\lambda^a \rangle^{1/2}
\approx [\big(\Gamma^{-1})^{aa}]^{1/2}.
\ee

The actual inner product, Eq.~(\ref{inner}), is formulated in the
frequency domain. For a white noise [i.e., $S_h(f) =$ constant],
the inner product is equivalent to $2 S_h^{-1}\sum_{\alpha=I}^{II}
\int_{-\infty}^{\infty} \, p_\alpha(t) q_\alpha(t) dt$,
by Parseval's theorem. Motivated by this formula, we shall adopt
the following approximate version of the inner product in calculating
the Fisher matrix:
First, we define the ``noise-weighted'' waveform
\be\label{replace}
\hat h_{\alpha}(t) \equiv \sum_n  h_{\alpha,n}(t)/S^{1/2}_h
\bigl(f_n(t)\bigr),
\ee
where (again) we take $f_n(t) = n\nu(t) + \dot{\tilde\gamma}(t)/\pi$.
Then we approximate the covariance matrix, Eq.~(\ref{sig}), as
\be\label{inner_approx}
\Gamma_{ab} = 2\big(\frac{3}{20}\big)\sum_{\alpha=I}^{II}\int_0^T{\partial_a
\hat h_{\alpha}(t) \partial_b \hat h_{\alpha}(t) dt} \, .
\ee
That is, we simply re-weight each harmonic by the square root of
the inverse spectral density of the noise, and thereafter treat the
noise as if it were white.

Finally, we must specify the (sky-averaged) noise spectral density $S_h(f)$
that we use in constructing the inner product.  For this, we use the noise model
given in Eq.~(30) of \cite{BC2}.  This model treats the noise as the
sum of instrumental noise and confusion noise from galactic and extragalactic
white dwarf binaries. All low-frequency binaries contribute additively to the
confusion noise, but the effect of the high-frequency galactic binaries
(which can all be detected and fitted out of the data) is to effectively reduce the
bandwidth for observing {\it other} sources (like EMRIs) by some factor,
or equivalently to {\it increase} the effective noise density $S_h(f)$ by that
same factor; see \cite{BC2} for more details.


\section{Results}

In this section we present our results for $\Delta\tilde Q$, the accuracy with which LISA can
determine the MBH's (dimensionless) quadrupole moment $\tilde Q\equiv Q/M^3$,
for a range of physical parameters.
This paper does not aim at a full, Monte Carlo-type exploration of $\Delta\tilde Q$ over the
entire $15$-dimensional parameter space.
Rather, we regard this paper as an initial exploration, aimed at determining the order
of magnitude of $\Delta\tilde Q$ for a few ``typical'' cases.  Based on this goal,
our survey strategy was as follows.  We first attempted to find values of the angular
parameters $(\theta_S,\phi_S,\theta_K,\phi_K,\lambda)$ that led to median results for
$\Delta\tilde Q$.  We then fixed these angles and explored the rest of parameter space.
We always evaluated the Fisher matrix at the value of $\tilde Q$
corresponding to pure Kerr geometry, i.e., $\tilde Q=-(S/M^2)^2$; thus, in effect,
we explored how accurately small deviations from the Kerr value could
be measured. In Sec.\ \ref{SecIIIA} we describe how we obtained ``representative'' values for
the angular parameters.  In Sec.\ \ref{SecIIIB} we present our results for $\Delta\tilde Q$, for
a variety of values of $M$, $\mu$, $S$, and $e_{\rm LSO}$.  Finally in Sec.\ \ref{SecIIIC} we present
checks on the accuracy and robustness of our numerical results.

\subsection{Choice of Representative Angles} \label{SecIIIA}

\begin{figure}[htb]
\includegraphics[width=10cm]{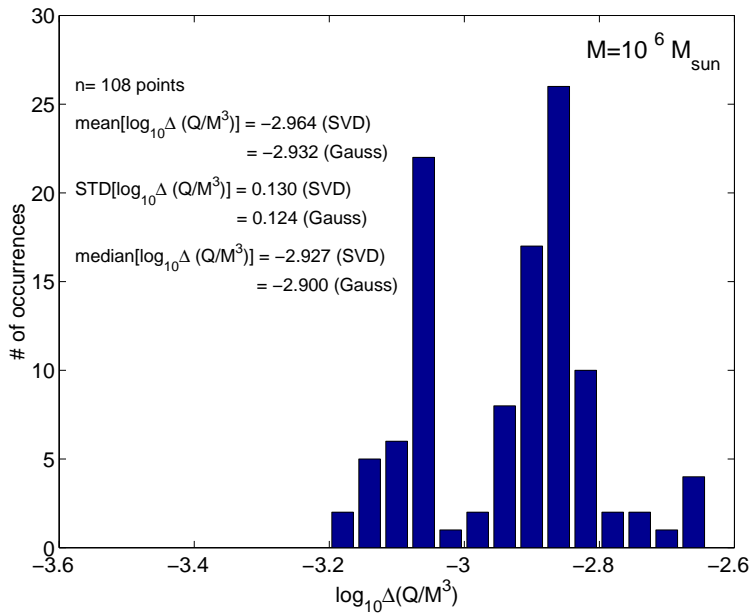}
\caption{\label{fig:median}
Distribution of $\Delta\tilde Q$ for a selection of
$(\theta_S,\phi_S,\theta_K,\phi_K,\lambda)$ values, and fixed values of
$(\mu,M,e_{\rm LSO})=(10 M_{\odot},10^6 M_{\odot},0.15)$.
The mean, standard deviation (STD), and median of the distribution are calculated
twice, for comparison, based on data from two different matrix-inversion methods:
singular-value decomposition (SVD) and Gauss-Jordan elimination (see discussion in the text).
We find that the choice $(\theta_S,\phi_S,\theta_K,\phi_K,\lambda)=
\left(\frac{2\pi}{3},\frac{5\pi}{3},\frac{\pi}{2},0,\frac{\pi}{3}\right)$
yields $\Delta\tilde Q$ very close to the median value.
The median values for the masses $M=10^{5.5}$ and $10^{6.5} M_{\odot}$ were obtained
in a similar manner, and are given in Eq.\ (\ref{median}).
The standard deviations in $\log_{10}\Delta\tilde Q$ were found to be $\sim 0.13$ and
$\sim 0.17$ for $M=10^{5.5}$ and $10^{6.5} M_{\odot}$, respectively.
}
\end{figure}

Here we describe our method for choosing one set of representative angles.
As we show below, $\Delta\tilde Q$ depends strongly on $M$, and more weakly
on $\mu$, $e_{\rm LSO}$, and $S$. We therefore fixed $\mu=10 M_{\odot}$,
$e_{\rm LSO}=0.15$, and $\tilde S=0.5$, and for each of the MBH masses
$M=10^{5.5} M_{\odot}$, $10^6 M_{\odot}$, and $10^{6.5} M_{\odot}$
explored all 108 different values of
$(\theta_S,\phi_S,\theta_K,\phi_K,\lambda)$ from within
$\theta_S\in\left(\frac{1}{6}\pi,\frac{1}{2}\pi,\frac{2}{3}\pi\right)$,
$\phi_S\in\left(0,\frac{2}{3}\pi,\frac{5}{3}\pi\right)$,
$\theta_K\in\left(\frac{1}{20}\pi,\frac{1}{2}\pi,\frac{3}{4}\pi\right)$,
$\phi_K\in\left(0,\frac{1}{2}\pi,\pi\right)$, and
$\lambda\in\left(\frac{1}{10},\frac{1}{3}\pi,\frac{2}{3}\pi\right)$.
We calculated $\Delta\tilde Q$ for each of these points and found the median
values of $\Delta\tilde Q$ for each of $M=10^{5.5},10^6,10^{6.5} M_{\odot}$.
The results for $M=10^6 M_{\odot}$ are histogrammed in Fig.~\ref{fig:median}.
The following choice of angles turns out to yield a value of $\Delta\tilde Q$
close to its median:
\begin{equation} \label{median}
(\theta_S,\phi_S,\theta_K,\phi_K,\lambda)=
\left\{ \begin{array}{ll}
\left(\frac{\pi}{2}, \frac{5\pi}{3},\frac{\pi}{2},0,\frac{\pi}{3}\right), & {\rm for\ } M=10^{5.5} M_{\odot},  \\
\left(\frac{2\pi}{3},\frac{5\pi}{3},\frac{\pi}{2},0,\frac{\pi}{3}\right), & {\rm for\ } M=10^6 M_{\odot}, \\
\left(\frac{\pi}{6},\frac{2\pi}{3},\frac{3\pi}{4},\pi,\frac{\pi}{3}\right), & {\rm for\ } M=10^{6.5} M_{\odot}.
\end{array} \right.
\end{equation}
We adopted these as our representative angles in surveying the $(\mu,S,e_{\rm LSO})$
space (for each of the different MBH values explored).

\subsection{LISA's measurement accuracy for $\tilde Q$}\label{SecIIIB}
In this subsection we present our results for $\Delta\tilde Q$, obtained by calculating the Fisher matrix and
using the leading-order estimate $(\Delta\tilde Q)^2 = (\Gamma^{-1}\big)^{\tilde{Q}\tilde{Q}}$.
All results are based on analyzing exactly one year of data, which we take to be the last year of inspiral,
and all are normalized to a total SNR of $100$.
Our reason for normalizing to SNR$ = 100$ is that the best tests of ``Kerr-ness'' will
naturally come from the strongest (generally closest) EMRI sources.  The detection threshold
for EMRIs, as limited by computational power, is estimated to be SNR$ \sim 30-35$~\cite{EMRIs}.
Assuming $\gtrsim 100$  EMRIs are detected above this
threshold, an estimated $\gtrsim 4$ EMRIs will be detected with SNR$ \gtrsim 100$.
Thus SNR$ =100$ corresponds to a relatively strong source, but we do expect to detect a
few sources that strong.  Of course, $\Delta\tilde Q$ roughly scales like
SNR$^{-1}$, so even for marginally detectable sources with SNR$ \sim 30-35$,
$\Delta\tilde Q$ will only be a factor $\sim 3$ larger.

\begin{figure}[htb]
\includegraphics[width=8.5cm]{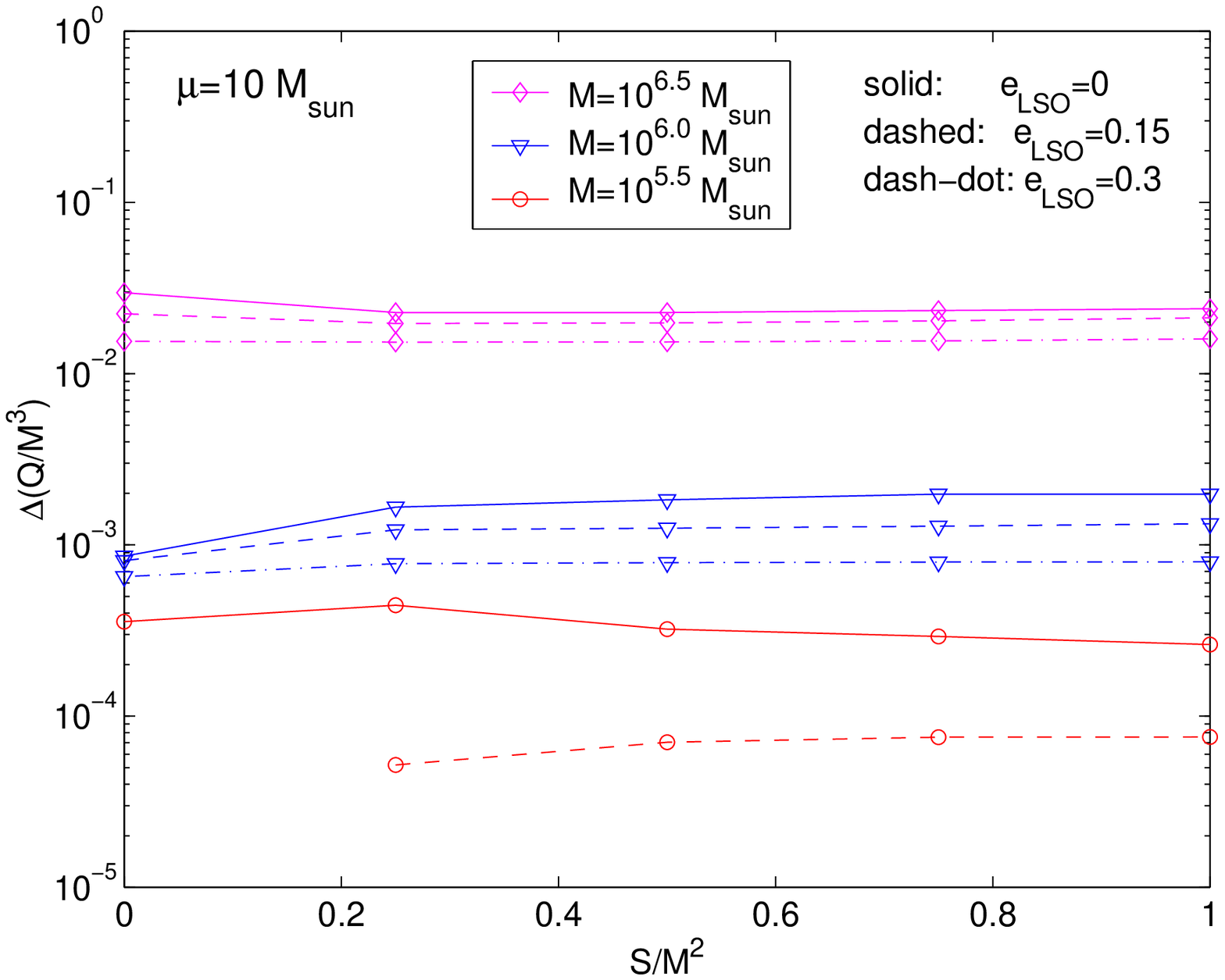}
\includegraphics[width=8.5cm]{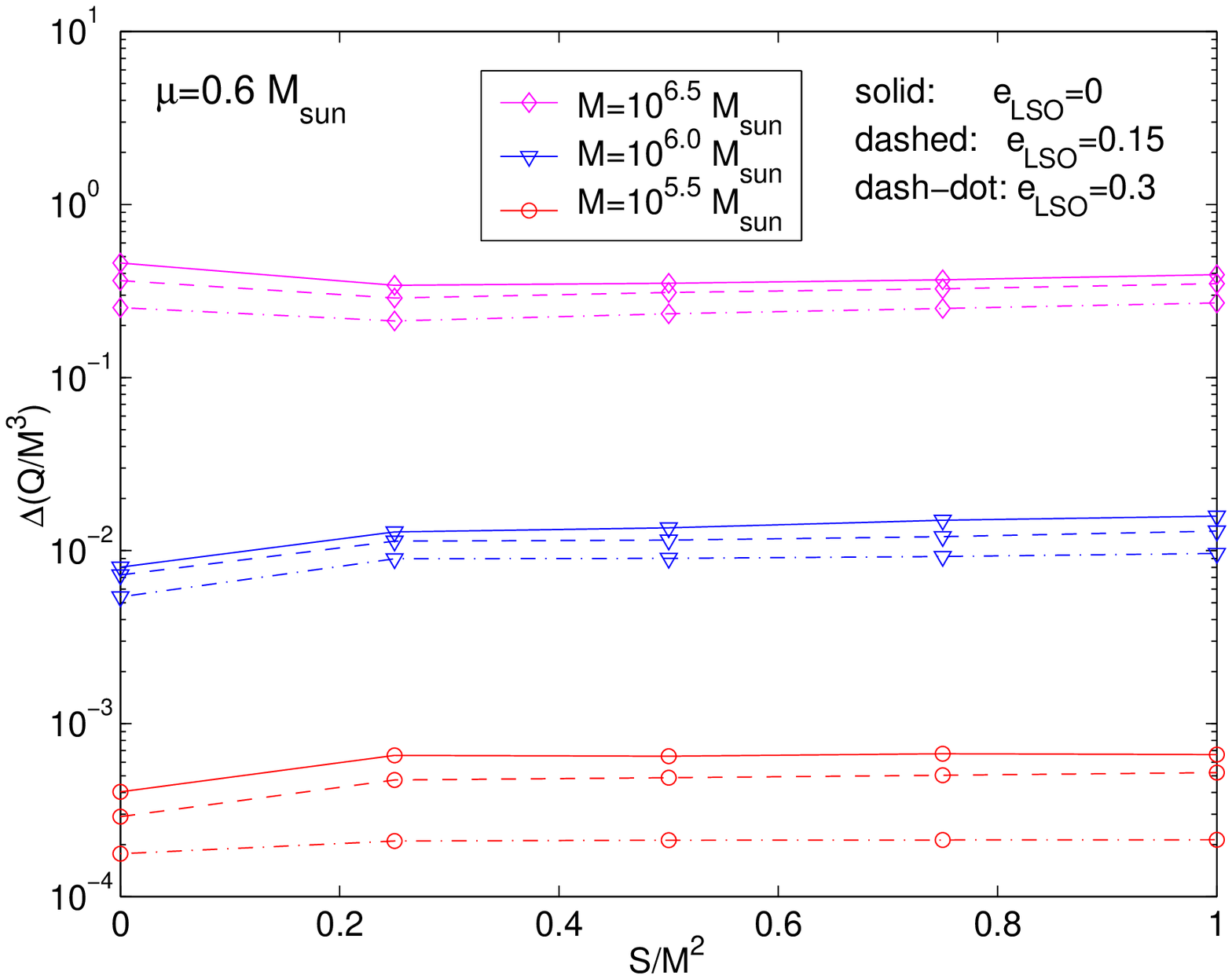}
\caption{\label{fig:Deltaq1}
Results for $\Delta\tilde Q$---the measurement accuracy in $\tilde Q\equiv Q/M^3$---for
a selection of values of $M$, $e_{\rm LSO}$,
and $\tilde S\equiv S/M^2$. Left and right panels display $\Delta\tilde Q$ for
$10 M_{\odot}$ and $0.6 M_{\odot}$ COs, respectively. Both plots are for one
year of data (the last year of inspiral), and are normalized to SNR$ = 100$.
[In the left panel we have discarded a few of the data points, corresponding
to low MBH mass with high LSO eccentricity: For these points the initial eccentricity
exceeds $0.7$, and our PN evolution model cannot be trusted (see discussion in the text).]
}
\end{figure}


Fig.~\ref{fig:Deltaq1}
presents our results for $\Delta\tilde Q$ for two fiducial CO masses ($10 M_{\odot}$ BH
and $0.6 M_{\odot}$ white dwarf), three fiducial MBH masses ($10^{5.5}$, $10^{6.0}$, and
$10^{6.5} M_{\odot}$), three fiducial values of $e_{\rm LSO}$ ($0$, $0.15$, and $0.3$),
and five fiducial values of $\tilde S$ ($0$, $0.25$, $0.5$, $0.75$, and $1$).
Again, all results in these figures are for the sets of ``representative'' angles
specified in Eq.\ (\ref{median}). We have not included results for
$(\mu,m,e_{\rm LSO})=(10M_{\odot},10^{5.5}M_{\odot},0.3)$
[and also missed out the point
$(\mu,m,e_{\rm LSO},\tilde S)=(10M_{\odot},10^{5.5}M_{\odot},0.3,0)$]
as for these parameter values the initial eccentricity becomes too high
($e\gtrsim 0.7$), and our evolution model fails to yield reliable results.
(The failure of our kludge model at very high eccentricities is discussed
in Appendix B of BC.)


The following general conclusions are evident from the above Figures.
For captured $10M_{\odot}$ BHs it should be possible to extract $\tilde Q$ to
within an accuracy $\Delta\tilde Q \sim 10^{-4}$--$10^{-2}$. The value of $\Delta\tilde Q$
varies strongly with $M$, being $\sim 100$ times larger for $M = 10^{6.5} M_{\odot}$
than for $M = 10^{5.5} M_{\odot}$.
As $\mu$ decreases from to $10$ to $0.6 M_{\odot}$ (with SNR assumed fixed),
$\Delta\tilde Q$ increases by a factor $\sim 10$. These patterns can be understood
qualitatively as follows. At fixed SNR, one expects $\Delta\tilde Q$, $\Delta ({\rm ln} M)$,
$\Delta ({\rm ln} \mu)$, and $\Delta\tilde S$ to all decrease with an increasing number
of cycles, and the latter scales roughly as $\nu_{\rm LSO}$ (for a fixed observation time),
or equivalently as $M^{-1}$. One also expects $\Delta\tilde Q$, $\Delta ({\rm ln} M)$,
$\Delta ({\rm ln} \mu)$, and $\Delta\tilde S$ to all decrease with an increasing ratio of
$f_{2,{\rm LSO}}/f_{2,{\rm yr}}$,
where $f_{2,{\rm yr}}$ is the frequency of the $n=2$ harmonic one year prior to plunge, and
$f_{2,{\rm LSO}}$
is that harmonic's frequency just prior to plunge.
That is because a larger ratio of final to initial frequencies provides more ``leverage''
for separating out post-Newtonian effects that enter
at different orders of the PN expansion parameter, $(\pi M f_2)^{1/3}$.
We can obtain an approximate expression for $f_{2,{\rm LSO}}/ f_{2,{\rm yr}}$
by simplifying to the case of $e=0$ orbits in Schwarzschild, so
$f_{2,{\rm LSO}}  = (6^{3/2}\pi M)^{-1}$, and by approximating $df_2/dt$  using
the lowest (nontrivial) order PN expression:
\begin{equation}
\frac{96}{5} \pi^{8/3} \mu M^{2/3} \int_0^{1 {\rm yr}} dt  =
\int_{f_{2,{\rm yr}}}^{f_{2,{\rm LSO}}} f^{-11/3}\, df \, ,
\end{equation}
which yields
\begin{equation}\label{fLSO}
\left(\frac{f_{2,{\rm LSO}}}{f_{2,{\rm yr}}} \right)^{8/3} = 1+ 2.53 \frac{\mu}{10 M_{\odot}}
\left(\frac{10^6 M_{\odot}}{M}\right)^2.
\end{equation}
Therefore, for the $\mu = 10 M_{\odot}$ case, as $M$ increases from $10^{5.5}$ to $10^{6.5} M_{\odot}$,
$f_{2,{\rm LSO}}/f_{2,{\rm yr}}$ decreases from $3.4$ to $1.09$.  Similarly, for $M=10^6 M_{\odot}$,
$f_{2,{\rm LSO}}/f_{2,{\rm yr}}$ decreases from $1.6$ to $1.05$ as $\mu$ decreases from
$10$ to $0.6 M_{\odot}$.

When one adds $\tilde Q$ to the list of parameters that must be extracted, the
accuracy with which all other parameters can be extracted {\it must} decrease. The magnitude of this
effect, in our case, is illustrated in Table~\ref{others}, where we list $\Delta(\ln\mu)$,
$\Delta(\ln M)$, and $\Delta\tilde S$ for our three fiducial $M$ values and $\mu= 10 M_{\odot}$.
The bold-faced entries are the error magnitudes for the $15$-dimensional parameter set including
$\tilde Q$, while the
plain-faced entries are for the $14$-dimensional set that excludes $\tilde Q$.
As expected, the errors when $\tilde Q$ is included
are always larger, by factors that range from $1.13$ to $54$.
Including $\tilde Q$ has the greatest impact on $\Delta\tilde S$.
This is also to be expected, since the terms $\propto \tilde Q$ and those $\propto \tilde S$ in
Eqs.\ (\ref{nudot}), (\ref{Gamdot}), and (\ref{alphadot}) have
a very similar scaling with frequency: the former are multiplied by $(2\pi M \nu)^{4/3}$
and the latter by $(2\pi M \nu)$; thus one expects errors in the two quantities to be
strongly correlated, and so also for $\tilde Q$ to have the strongest effect on $\tilde S$.
The effect of the extra parameter $\tilde Q$ on the size of $\Delta \tilde S$ is largest for
high $M$, since the smaller the ratio $f_{2,{\rm LSO}}/f_{2,{\rm yr}}$, the more degenerate these
parameters become.

\begin{table}
\begin{tabular}{|c|c|c|c|} \hline
$M$ & $\Delta(\ln\mu)$ & $\Delta(\ln M)$ & $\Delta\tilde S$
\\\hline \multirow{2}{20mm}{\centering $10^{5.5} M_{\odot}$} &
$\mathbf{2.9\times 10^{-6}}$ & $\mathbf{6.9\times 10^{-7}}$ &
$\mathbf{8.3\times 10^{-5}}$ \\ & $1.8\times 10^{-6}$ & $4.7\times
10^{-7}$ & $1.7\times 10^{-6}$
\\\hline
\multirow{2}{20mm}{\centering $10^{6} M_{\odot}$}
& $\mathbf{5.2\times 10^{-5}}$ & $\mathbf{1.6\times 10^{-6}}$ & $\mathbf{1.6\times 10^{-4}}$ \\
& $2.5\times 10^{-5}$ & $1.1\times 10^{-6}$ & $3.6\times 10^{-6}$
\\\hline
\multirow{2}{20mm}{\centering $10^{6.5} M_{\odot}$}
& $\mathbf{8.8\times 10^{-4}}$ & $\mathbf{2.6\times 10^{-6}}$ & $\mathbf{2.6\times 10^{-3}}$ \\
& $5.9\times 10^{-4}$ & $2.3\times 10^{-6}$ & $4.8\times 10^{-5}$
\\\hline
\end{tabular}
\caption{\protect\footnotesize
Bold-face data in this table show the measurement accuracy of
$\ln\mu$, $\ln M$, and $\tilde S$ when $\tilde Q$ is included as a search parameter.
(Here We have taken $\mu=10 M_{\odot}$, $e_{\rm LSO}=0.15$, $S=M^2$, and a few
values of $M$, which corresponds to three of the data points in Fig. 2 above.)
For comparison, plain-face data show the measurement accuracy of these
parameters for the same points in parameter space, but without searching over $\tilde Q$,
i.e., assuming a Kerr black hole: $\tilde Q = -(S/M^2)^2$.
}\label{others}
\end{table}

\subsection{Inversion checks and test of robustness} \label{SecIIIC}
The  Fisher matrices we compute typically have a very large condition number
(the ratio of the largest to smallest eigenvalues).
For the numerical computation of a matrix inverse to be reliable, using double precision
arithmetic, it is generally sufficient (but not necessary!) that the matrix's
condition number be $\lesssim 10^{14}$~\cite{NumRec}.~\footnote{As an example
of a matrix that fails this condition but that
is easy to invert numerically, consider the $2\times 2$ matrix with diagonal elements $1.0$ and $10^{-30}$, with
zeroes for the off-diagonal elements.}  Some of the Fisher matrices we calculate have condition numbers as
large as $\sim 10^{22}$.
This is cause for some concern when we numerically invert the Fisher matrices to find $\Delta\tilde Q$;
specifically, one might worry that small errors in the Fisher matrix elements
(e.g., due to round off)
could lead to large errors in the matrix inverse.  Therefore we have performed several checks that
our results are reliable.

\begin{figure}[htb]
\includegraphics[width=10cm]{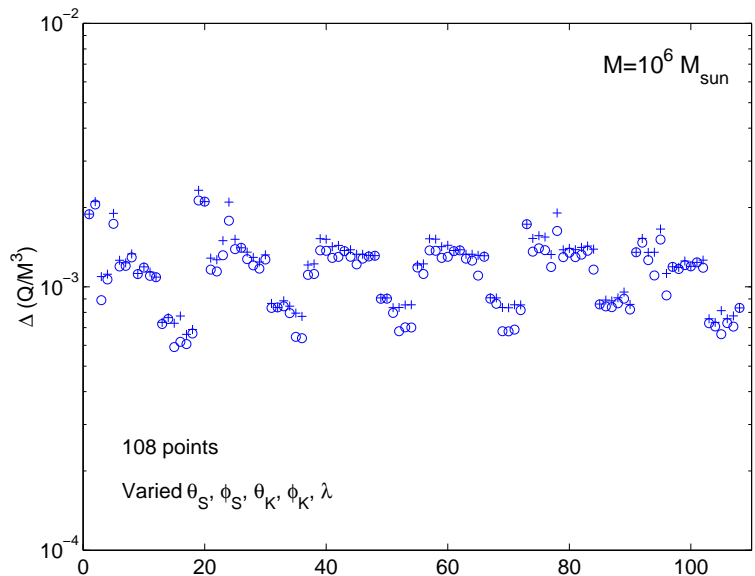}
\caption{\label{fig:inversions}
Comparison of the values of $\Delta\tilde Q$ obtained using two different matrix inversion
methods.  The 108 sample points are the same ones displayed in Fig.\ \ref{fig:median};
i.e., they correspond to 108 different values for
$(\theta_S,\phi_S,\theta_K,\phi_K,\lambda)$, and fixed values of
$(\mu,M,e_{\rm LSO}, \tilde Q)$.  Circles are results using singular-value decomposition, while
pluses correspond to inversion of the same matrices using Gauss-Jordan elimination.
Generally the two results lie close together, with deviations smaller
than $\sim 20 \%$ in all cases.  This is sufficient agreement for our purpose.
}
\end{figure}

First, we have used several different matrix-inversion routines, and checked
that the results from different routines are always fairly close together.
This is illustrated in Fig.~\ref{fig:inversions}, where for the $108$ matrix inversions
used to generate Fig.\ \ref{fig:median}, we plot the values of $\Delta\tilde Q$
derived from two different methods: (i) {\it Gauss-Jordan elimination} and
(ii) {\it singular-value decomposition} \cite{NumRec}.
The results from the two different inversion routines agree to within
less than  $\sim 20 \%$ in all cases.
Additional evidence that our numerical inverses are numerically stable come
from the plots in Fig.\ \ref{fig:Deltaq1}, which show that $\Delta\tilde Q$
depends in a rather smooth and consistent way on changes in $\tilde S$ and
$e_{\rm LSO}$; i.e, the matrix inverse is
clearly {\it not} wildly sensitive to changes in source parameters.
Thirdly, if we recondition our matrices by essentially re-scaling all the variables,
so that diagonal elements of the new Fisher matrix are all unity, then a matrix with
an original condition number of $\sim 10^{22}$ typically has a new condition
number as low as $\sim 10^{12}$.

It should be pointed out that the error in some of the orbital phases (those associated
with the smallest singular values) {\it are} strongly dependent on the inversion
method, and our code does not produce reliable answers for these (apart from telling
us that these parameters are strongly degenerate). Fortunately, the parameter $\tilde Q$
is among those for which the inversion gives robust answers.

A related worry is that
our results for $\Delta\tilde Q$ might depend strongly on the exact signal model, so that $\Delta\tilde Q$ might
be very different for our ``analytic kludge''  waveforms than for true, general relativistic waveforms.
We now present evidence that, to the contrary, our general conclusions seem rather robust.
Roughly speaking, our ``analytic kludge'' waveforms differ from the true ones due
mainly to all the higher-order PN terms that are missing from the former.
(Our treatment of the LISA response is also approximate, but our low-frequency approximation
here seems much less likely to significantly bias the final results.)
We can get some sense
of the importance of these missing terms for the final conclusion, by comparing our results
with a ``lower-order'' or ``dumbed-down'' version  of the analytic kludge.
Specifically, the dumbed-down
version just neglects all the correction terms proportional to $(2\pi M \nu)^{2/3}$ in
Eqs.\ (\ref{nudot})--(\ref{alphadot}) . That is, we simply drop the second line in
Eq.(5), the second term in square
brackets in the first line of Eq.(6), and the second line of Eq.(7).  The remaining equations still
have the property that the major qualitative features of the orbits (perihelion precession,
Lense-Thirring precession, and inspiral) are all present, and that at least the lowest-order PN term
corresponding to each effect is included.

\begin{figure}[htb]
\includegraphics[width=10cm]{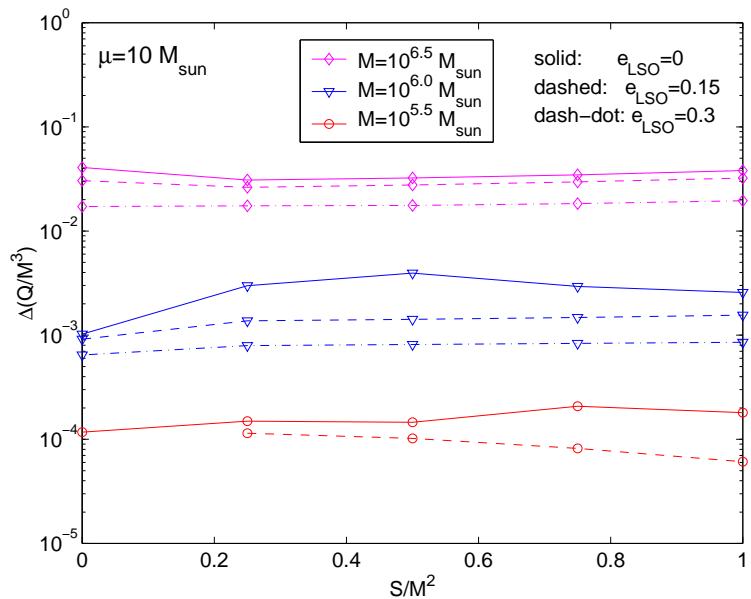}
\caption{\label{fig:robust}
$\Delta\tilde Q$ for BH inspirals, for a lower-order version of
our ``kludge'' evolution equations (given explicitly in the text).
Comparison of this Figure with the left panel of
Fig.\ \ref{fig:Deltaq1}   provides a check on the
robustness of our results.
}
\end{figure}

Our results for $\Delta\tilde Q$, using this ``dumbed-down'' version of the analytic
kludge evolution equations, are demonstrated in Fig.\ \ref{fig:robust} for the
case $\mu = 10 M_{\odot}$. The values of $\Delta\tilde Q$
remain the same to within a factor $\sim 2$ (compare with the left panel of
Fig.\ \ref{fig:Deltaq1}).  We take these results as indication that
our estimates for $\Delta\tilde Q$ are accurate to better than an order
magnitude---perhaps even to within a factor $\sim 3$.

\section{Conclusions and Discussion}

Our main conclusion is that, for reasonably strong EMRI sources (SNR$ \sim 100$), it should
be possible to extract the quadrupole moment $Q$ of the central massive object, independently
of its mass and spin, typically to within $\Delta \tilde Q \sim 10^{-4}$--$10^{-2}$.
Thus for each detected EMRI, LISA will have the opportunity to perform a rather precise
and nontrivial check of the ``Kerr-ness'' of the central massive object.
For fixed SNR, smaller values of $M$ will provide much more accurate measurements of
the MBH's quadrupole moment. While the particular source orientation and sky direction
can cause $\Delta\tilde Q$ to vary by up to a factor $~4$,
$\Delta\tilde Q$ seems to depend rather weakly on
$e_{\rm LSO}$ and on the value of the spin $S$.


An obvious caveat is that
the values of $\Delta\tilde Q$ derived here
represent the magnitude of the error due to noise alone;
any errors in the construction of theoretical waveforms for given parameter values would lead to additional, systematic
errors in $Q$.
Another caveat is that even if the best-fit values of $Q$ were substantially off-Kerr in a few sources,
this could just mean that those sources were not
sufficiently ``clean'' two-body systems, e.g., because of the influence of an accretion
disk or because of interactions with a third body.

Finally, it might be interesting to extend our analysis to consider independent estimates of the first $N$ multipole moments
(for $N = 3, 4, 5, \ldots$).  Indeed this is what Ryan~\cite{ryan97I} did, within his highly simplified model.  However we have not yet attempted
this.


\section{acknowledgments}
We thank for Kip Thorne for encouraging us to work on this problem.
We thank Neil Cornish for pointing out
a typographical error in Eq.\ (33) of BC.
L.B.'s work was supported by PPARC through grant number PP/D001110/1.
L.B. also gratefully acknowledges financial support from the Nuffield
Foundation, and thanks Kostas Glampedakis for helpful discussions.
C.C.'s work was carried out at the Jet Propulsion Laboratory, California Institute of Technology,
and was sponsored by the National Aeronautics and Space Administration.


\end{document}